\documentclass{AUJarticle}
\usepackage[cmex10]{amsmath}
\usepackage[utf8x]{inputenc}
\usepackage[nocompress]{cite}
\usepackage{graphicx, multirow, booktabs, color, listings}
\usepackage{balance}
\usepackage{subcaption}
\usepackage{xcolor}
\usepackage[export]{adjustbox}
\usepackage{pdfpages}
\usepackage{multirow}
\usepackage{hyperref}

\graphicspath{{figures/}}

\begin{document}

\title{SysML Modeling of Digital Twins for Renewable Energy Communities}

\addauthor{Mohammad Samadi and Lu\'{i}s Miguel Pinho}
{Polytechnic Institute of Porto \& INESC TEC, Portugal}
{\{mmasa,lmp\}@isep.ipp.pt}

\addauthor{Andrey Sadovykh}
{Softeam, Paris, France}
{andrey.sadovykh@softeam.fr}

\addauthor{Gabriela Lucas}
{Cleanwatts Digital, Coimbra, Portugal}
{glucas@cleanwattsdigital.com}

\shortauthor{M. Samadi, L. M. Pinho, A. Sadovykh, G. Lucas}
\shorttitle{SysML Modeling of Digital Twins for RECs}

\thispagestyle{plain}

\maketitle

\begin{abstract}
Renewable Energy Communities (RECs) are emerging as a key organizational model for local and global sharing of renewable generation, storage, and flexible loads. Engineering Digital Twins of RECs is made difficult by the heterogeneity of devices, contracts, and runtime data involved. In this paper, we take a first step toward a Model-Based Systems Engineering (MBSE) workflow for REC's Digital Twins. Starting from an industrially-validated REC domain model
, we re-express a representative \emph{house} subset in SysML using the open-source Modelio tool, yielding two Block Definition Diagrams -- a device taxonomy and a community organizational view. We then discuss four semantic gaps that plain SysML leaves open and sketch how the SAREF4ENER ontology could be imported as a reference package to close them. Combining SysML with SAREF-based semantics for smart-energy Digital Twins remains largely unexplored, and we position this paper as a first step along that line.

\vspace{0.5em}
\end{abstract}

\section{Introduction}
Renewable Energy Communities (RECs) are an emerging organizational form in which local consumers and prosumers -- households, small businesses, and public facilities -- pool their photovoltaic (PV) generation, storage, electric vehicles (EVs), and flexible loads to share energy and participate in grid-level demand response. Managing a REC in practice requires simultaneously monitoring dozens of heterogeneous Internet of Things (IoT) devices, honoring contractual constraints between members, and meeting grid and regulatory limits. The complexity of this problem has made Digital Twins (DTs) a natural abstraction: a live (real-time), virtual counterpart of each physical device that aggregates measurements, exposes controllable set-points, and can be used for optimization, what-if analysis, and predictive maintenance~\cite{alshetwi2025dt
,abdelrahman2025definition}.

Implementing such DTs in the REC context, however, is non-trivial. The relevant artifacts -- device APIs, ontologies, flexibility profiles, and contracts -- live at very different levels of abstraction and are usually produced by different stakeholders. Without an explicit and shared engineering model, the system quickly becomes a point-to-point integration effort. Model-Based Systems Engineering (MBSE)~\cite{pessoa2023mbse} and, more recently, Model-Driven Digital Twin Engineering~\cite{lehner2025mde
} have been proposed to tame this complexity, but most published works target manufacturing or generic cyber-physical systems, and therefore the REC domain is still underexplored~\cite{alshetwi2025dt
}.

In this paper, we take a first, pragmatic step towards such a workflow. Starting from an industrially-validated REC domain model
, we re-express a representative subset of it as plain Systems Modeling Language (SysML) using the open-source Modelio tool~\cite{modelio}. The result is a pair of Block Definition Diagrams (BDDs) -- a device taxonomy and a REC community view -- which together capture the structural skeleton of a REC Digital Twin. We use a compact \emph{household} scenario as the running example. Building on this model, we identify four semantic gaps that plain SysML leaves open, and we sketch how the SAREF4ENER ontology could be imported as a reference package to close them. The mapping itself remains future work: our contribution here is the SysML model plus an explicit statement of the gap it does not yet close. It is worth mentioning that the main objective is also to provide a Digital Twin (DT) engineering workflow, but in this paper, we focus primarily on modeling the organizational layer and the energy assets (i.e., devices) needed to build a REC Digital Twin. The main contributions of this work can be represented as folows:
\begin{itemize}
  \item Modeling the organizational layer and the energy assets using Modelio based on a real-world use case.
  \item Identification of four semantic gaps in SysML and introduction of the SAREF4ENER ontology as a reference to fill them.
  \item Addressing a DT engineering workflow for RECs with different IoT energy devices.
\end{itemize}

The paper is organized as follows. Section~\ref{sec:background} briefly presents the specifications of REC Digital Twins, SysML modeling, and an ontological approach to managing physical devices with different configurations in DTs. Section~\ref{sec:approach} describes the proposed modeling using the REC domain model and SysML modeling in Modelio. Section~\ref{sec:scenario} illustrates an example of a household scenario, including the organizational layer of REC and the specifications of different energy assets. Sections \ref{sec:discussion} and \ref{related_work} address the gaps in SysML modeling and the related work for REC DTs, respectively. Finally, Section \ref{conclusion} concludes the paper.

\section{Background}
\label{sec:background}
A Renewable Energy Community (REC) is a legally recognized association of local prosumers and consumers that share renewable electricity -- typically PV -- possibly supplemented with storage, EVs, and other devices. From an information-system perspective, a REC is a distributed cyber-physical system that continuously exchanges measurements and set-points with the grid and the member devices, under import/export limits and internal pricing rules. Recent surveys confirm DTs as a suitable abstraction to manage this complexity~\cite{alshetwi2025dt
}, with a Digital Twin understood throughout this paper as a live, bidirectionally synchronized virtual counterpart of a physical asset~\cite{abdelrahman2025definition}.

The engineering of such DTs benefits from established MBSE languages and tools. The SysML extends Unified Modeling Language (UML) with blocks, ports, flow specifications, activities, parametric diagrams, and requirements traceability, and has been repeatedly proposed as a structural backbone for DT engineering~\cite{wilking2022sysml4dt,pessoa2023mbse,ferko2025aas}. We build our model using \emph{Modelio}~\cite{modelio}, an open-source MBSE workbench that supports SysML and custom profiles, as well as Jython-based automation for diagram and artifact generation -- a combination that keeps the model extensible and reproducible.

Structural models alone, however, are not sufficient to guarantee semantic interoperability between stakeholders of heterogeneous smart-energy. The Smart Appliances REFerence (SAREF) ontology family, originally introduced by Daniele et al.~\cite{daniele2016interoperability} and standardized at ETSI\footnote{ETSI stands for the European Telecommunications Standards Institute and is a recognized organization for standardization.}, provides a shared vocabulary for smart appliances. Its energy extension, called SAREF for Energy (SAREF4ENER), refines this vocabulary with device categories, such as \texttt{PowerGenerator}, \texttt{LoadDevice}, or \texttt{Storage}, and with a family of flexibility profiles (\texttt{PowerProfile}, \texttt{PowerEnvelopeProfile}, \texttt{DemandDrivenProfile}, \texttt{FillRateProfile}) designed for demand-response interactions. Complementary DT runtimes, such as Eclipse~Ditto~\cite{kherbache2022ditto
}, provide a lightweight JSON-based \emph{Thing} model and MQTT/HTTP connectivity to close the loop with physical devices. Together, SysML, SAREF4ENER, and Eclipse~Ditto cover the design-, semantic- and runtime layers that a REC Digital Twin needs.

Ontologies are essential in the energy sector (e.g., in RECs) to manage the vast amounts of data contributed by systems and stakeholders with diverse configurations, allowing seamless integration and communications between different platforms. An ontology can be defined as a formal specification of a shared conceptualization, which organizes properties, concepts, and relationships into machine-readable models. SAREF4ENER stands out as an ontology for the energy sector, as it improves interoperability in energy systems and smart appliances, increases grid stability, integrates renewable energy sources, and improves overall energy efficiency. Compared to other ontologies, it focuses on energy management systems, offers high interoperability within energy systems, provides less integration complexity, and contains energy-specific data~\cite{chy2024design}.

\section{Proposed Modeling}
\label{sec:approach}
Engineering DTs for RECs involves heterogeneous concerns -- devices, contracts, semantics, and real-time data. MBSE offers a principled way to integrate them through formal models~\cite{
lehner2025mde}. We follow a three-layer approach: (i) a REC domain model
, (ii) its expression in SysML using Modelio, and (iii) a planned semantic bridge to SAREF4ENER, discussed in Section~\ref{sec:discussion}.

\subsection{
REC Domain Model}
A production-level UML domain model is proposed for a REC, structured around contexts bounded by Domain-Driven Design. It defines (a) a \emph{Community \& Infrastructure} layer with organizational entities -- REC, Site, User, Energy Contract, Membership, (b) an \emph{Entity (Device) Aggregate} with a taxonomy of eleven energy asset types -- PV Panel, Battery, EV, EV Charger, Heat Pump, Heater, Water Pump, Grid Meter, Micro Wind Turbine, Home Appliance, Non-Shiftable Load -- each characterized by static properties (nominal power, efficiency, SOC bounds), sensor readings, and actuation commands, and (c) a decomposition into \emph{Bounded Contexts} (Telemetry \& Time Series, Optimization \& Control, Access Control, Flexibility \& Participation). The model has been validated against real hardware (Shelly EM clamp meters exposed via a REST API), making it a concrete reference for the REC domain, rather than an abstract taxonomy.

\subsection{SysML Expression in Modelio}
To bring this domain model into an MBSE workflow, we re-express a representative subset of the model in SysML BDDs using the open-source Modelio tool~\cite{modelio}. We provide two complementary views.

\begin{figure}[t]
  \centering
  \includegraphics[width=0.95\columnwidth]{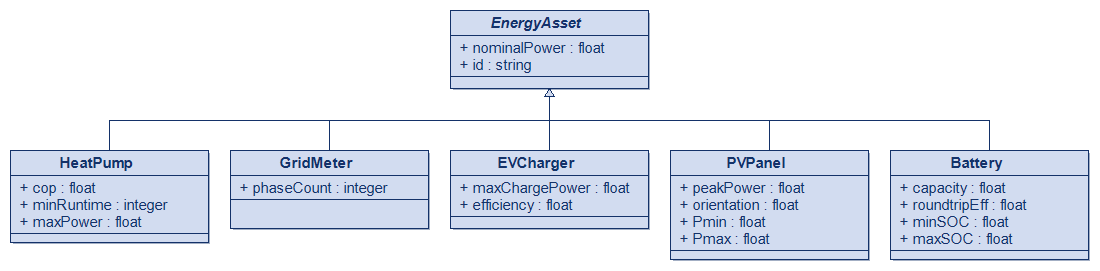}
  \caption{SysML BDD of the household. Five concrete blocks -- \texttt{HeatPump}, \texttt{GridMeter}, \texttt{EVCharger}, \texttt{PVPanel}, and \texttt{Battery} -- specialize a common abstract \texttt{EnergyAsset}, where the attributes use standard UML primitive types.}
  \label{fig:ada_bdd_devices}
\end{figure}

\emph{Device taxonomy} (Figure~\ref{fig:ada_bdd_devices}). A compact \emph{household
} view encompassing five representative device blocks -- \texttt{HeatPump}, \texttt{GridMeter}, \texttt{EVCharger}, \texttt{PVPanel}, and \texttt{Battery} -- all specialized in a common \texttt{EnergyAsset} block, carrying the shared attributes (\texttt{nominalPower}, \texttt{id}). Each concrete block adds its own typed value properties -- e.g., \texttt{capacity}, \texttt{roundtripEff}, \texttt{minSOC}, and \texttt{maxSOC} for Battery; \texttt{cop}, \texttt{minRuntime}, and \texttt{maxPower} for HeatPump. The types are standard UML primitives (\texttt{float}, \texttt{integer}, \texttt{string}, and \texttt{date}), keeping the diagram tool-independent.

\begin{figure}[t]
  \centering
  \includegraphics[width=0.95\columnwidth]{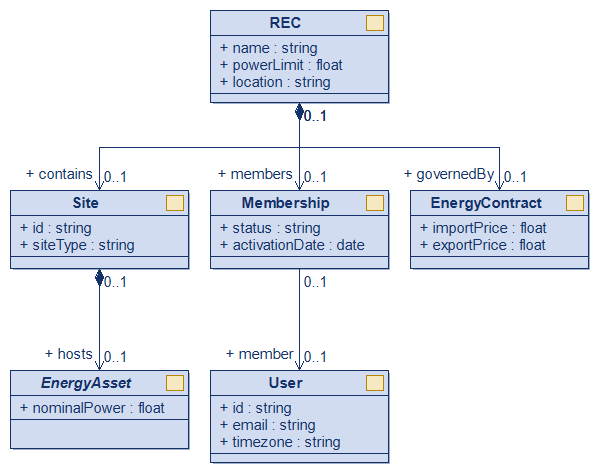}
  \caption{SysML BDD of the REC organizational layer: a \texttt{REC} aggregates \texttt{Site}s, \texttt{Membership}s and \texttt{EnergyContract}s; \texttt{Site}s host \texttt{EnergyAsset}s (cf. Figure~\ref{fig:ada_bdd_devices}), and \texttt{User}s are linked through memberships.}
  \label{fig:ada_bdd_rec}
\end{figure}

\emph{REC community model} (Figure~\ref{fig:ada_bdd_rec}). A second BDD captures the organizational layer around the devices: a \texttt{REC} contains \texttt{Site}s, \texttt{Membership}s (linking \texttt{User}s) and an \texttt{EnergyContract}. Each site hosts one or more \texttt{EnergyAsset}s. This view exposes what a purely device-centric DT typically omits -- the contractual and user layer that any REC deployment must take account for.

These compact diagrams do not replace 
the complete proposed UML catalogue. In fact, they illustrate how the 
abstractions -- entity hierarchy, site/membership decomposition, and user contracts -- map seamlessly into SysML blocks and associations. The same transformation scales to the other device types and bounded contexts.

\section{Household Scenario}
\label{sec:scenario}
To illustrate the workflow, we consider a single household that is a member of a REC and owns: a 5\,kWp rooftop PV system, a 10\,kWh lithium battery, an 11\,kW EV charger, and an air-to-water heat pump for heating and domestic hot water. The household pays time-of-use tariffs and participates in a demand-response program with the REC aggregator.

In the SysML model (Figure~\ref{fig:ada_bdd_devices}), each of these assets is a block with nominal properties, where the household belongs to a \texttt{Site} that is itself part of a \texttt{REC} governed by an \texttt{EnergyContract} (Figure~\ref{fig:ada_bdd_rec}). This skeleton is sufficient to map the real devices (Shelly EM clamps, inverter, wallbox, and heat-pump controller) into model-level blocks and to reason about aggregation, contracts, and membership at design time. What the skeleton does \emph{not} yet carry -- standardized semantics, flexibility profiles, and an active runtime link -- is the subject of Section~\ref{sec:discussion}.

\begin{figure}[t]
  \centering
  \includegraphics[width=0.95\columnwidth]{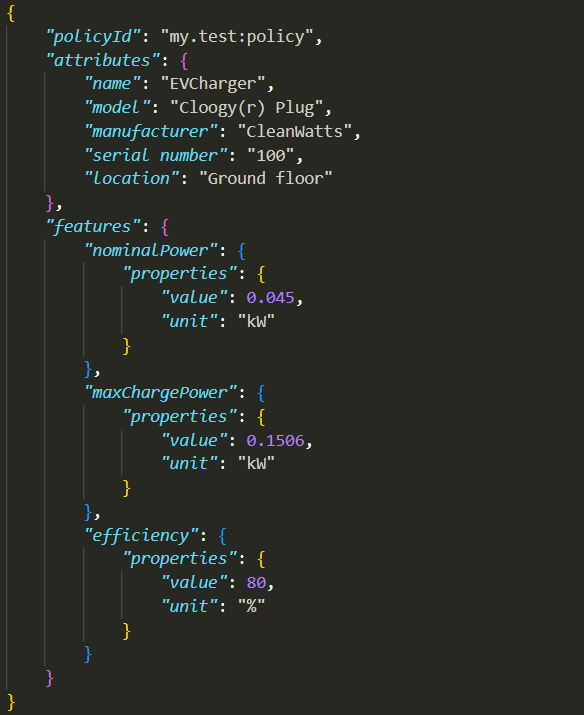}
  \caption{Representation of the \texttt{EVCharger} device in JSON format.}
  \label{fig:evcharger}
\end{figure}

Each device can be represented in a JSON format/structure (i.e., metamodel) in the implementation phase. This structure contains various features and properties defined for the device, which can be used to manage things in Eclipse~Ditto. Figure~\ref{fig:evcharger} shows an example, defining the specifications of \texttt{EVCharger} with three features: \texttt{nominalPower}, \texttt{maxChargePower}, and \texttt{efficiency}. Each feature is also represented by two properties: value and unit.

SysML~v2 improves upon the limitations of SysML~v1.6 by introducing a new metamodel, textual notation, enhanced interoperability, and extended features for representing complex system. It enhances interoperability across engineering domains through textual notation and models, enabling bidirectional data exchange, rapid access to current process information, and informed decision-making based on objective assessments~\cite{grunenwald2025leveraging}. Therefore, we have provided the metamodel for each device specification in a JSON structure to simplify the model and offer structural interoperability with different engineering tools. This can facilitate bidirectional data exchange with other platforms, system engineers, and managers, allowing easy access to up-to-date information on the current stages of the process and the system state.

There are also several Digital Twin interoperability standards, such as Asset Administration Shell (AAS), Digital Twins Definition Language (DTDL), and Next Generation Service Interface with Linked Data (NGSI-LD), described below.

AAS is an open metamodel framework created specifically for smart manufacturing and Industry 4.0. It serves as a standardized Digital Twin for industrial production lines, robots, and manufacturing equipment~\cite{sakurada2025ten}. AAS is not the primary architecture used in Cleanwatts devices; instead, IoT \& cloud architecture is used in these devices. However, SysML2AAS will be made available in future work, automating the transformation of SysML v2 models into AAS representations and facilitating compatibility with Industry 4.0 systems (e.g., BaSyX). It links engineering design models (e.g., microgrid topologies and state machines) to AAS submodels, enabling traceable requirements decomposition and a clear, auditable design-to-deployment connection.

Furthermore, properties, behaviors, and relationships between entities in a Digital Twin environment can be described using DTDL, an open-source modeling language based on JSON for Linked Data (JSON-LD). It serves as a standard model for Digital Twins, facilitating the integration of IoT systems and devices across different platforms~\cite{schmidt2023dtdl_aas}. DTDL is generally used in our Digital Twins model for RECs, as device specifications and configurations are represented in JSON/YAML format.

In addition, ETSI publishes the open, standard information model and API known as Next Generation Service Interface with Linked Data (NGSI-LD). It is provided to create, exchange, and query real-time data for Digital Twins, facilitating communication between various virtual and physical systems~\cite{gal2023industry}. This standard is also used in our proposed framework for Digital Twin RECS, as we have used REST API to obtain real-time data from Cleanwatts living labs via JSON format, sharing data between physical and virtual systems, as well as the system can also be represented using a knowledge graph.

\begin{figure*}[t]
  \centering
  \includegraphics[width=1.9\columnwidth]{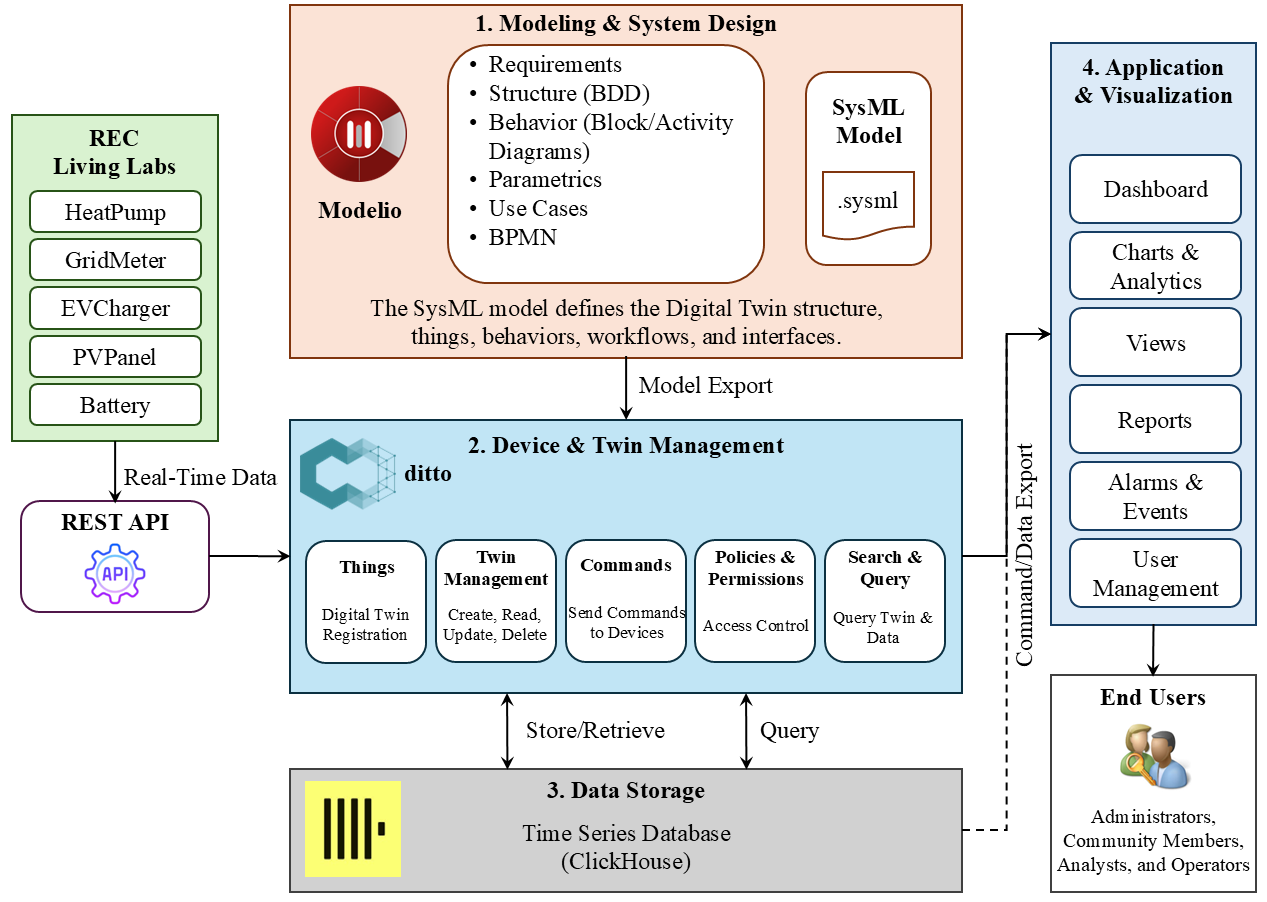}
  \caption{Digital Twin architecture for RECs, where the entire system is controlled through multiple modules.}
  \label{fig:Architecture}
\end{figure*}

Figure~\ref{fig:Architecture} shows a Digital Twin architecture for RECs, illustrating the role of Eclipse~Ditto beyond future integration. The processes for managing the DT RECs can be represented as follows. First, the system model is designed using Modelio to define structure (BDD), things, behaviors (block/activity diagrams), workflows (BPMN), and interfaces (parametrics). The model is then exported to Ditto for device and DT management. Afterwards, real-time data is obtained from the Cleanwatts living labs via the REST API, and Ditto is updated based on this data. The database is also updated according to the given real-time data. The application and visualization module is used to display various system information through charts, views, reports, etc. The system also handles permission control and end-user management. It is worth noting that the modeling part is performed in this paper, while the other parts will be addressed in the future integration.

\section{Discussion: Gaps and the SAREF Bridge}
\label{sec:discussion}
A SysML model captures \emph{structure} -- blocks, attributes, and associations -- but does not, by itself, carry \emph{semantics} understandable to third-party energy management systems. By inspecting the model in Section~\ref{sec:approach} against the needs of a REC Digital Twin, we identify four gaps, as described in Table~\ref{tab:gaps}.

\begin{table}[t]
\centering
\caption{Semantic gaps in plain SysML for REC modeling.}
\label{tab:gaps}
\small
\begin{tabular}{cp{2.6cm}p{3.2cm}}
\toprule
 & \textbf{Gap} & \textbf{Consequence} \\
\midrule
G1 & Plain classes, no ontology anchor     & No cross-system recognition of ``Battery'' as \texttt{s4ener:Storage} \\
G2 & Free-text sensor/actuation labels     & Same concept named differently across models \\
G3 & No flexibility / DR model             & Cannot express REC contractual flexibility \\
G4 & No runtime link                       & Design-time only, disconnected from the live state \\
\bottomrule
\end{tabular}
\end{table}

The SAREF ontology family and its SAREF4ENER extension~\cite{daniele2016interoperability,lilliu2025flexoffers} address G1--G3, providing a shared vocabulary for devices, measurements, commands, and flexibility profiles. Recent work extends SAREF4ENER with FlexOffers~\cite{lilliu2025flexoffers} and evaluates alternative smart energy ontologies~\cite{haghgoo2020sargon,ivanova2025unified}. Table~\ref{tab:saref_mapping} sketches the mapping from the blocks of Figure~\ref{fig:ada_bdd_devices} to the corresponding SAREF4ENER concepts and flexibility profiles.

\begin{table}[t]
\centering
\caption{Sketched mapping from SysML blocks to SAREF4ENER concepts.}
\label{tab:saref_mapping}
\small
\begin{tabular}{lll}
\toprule
\textbf{SysML Block} & \textbf{SAREF / SAREF4ENER Class} & \textbf{Profile} \\
\midrule
PVPanel    & \texttt{s4ener:PowerGenerator} & PowerEnvelope \\
Battery    & \texttt{s4ener:Storage}        & FillRate \\
EVCharger  & \texttt{s4ener:LoadDevice}     & Power \\
HeatPump   & \texttt{s4ener:LoadDevice}     & DemandDriven \\
GridMeter  & \texttt{saref:Meter}           & IncentiveTable \\
\bottomrule
\end{tabular}
\end{table}

Rather than encoding this mapping through heavy custom stereotypes (which would tie the model to a specific SysML profile dialect), we propose to realize it by importing SAREF4ENER as a dedicated \emph{reference package} in the same Modelio project. Consequently, each SAREF4ENER class becomes a plain reference block, and the REC device blocks declare typed dependencies or specializations related to them. This keeps the diagrams of Section~\ref{sec:approach} free of annotation clutter, while making the semantic anchor explicit and traceable in the model. Gap G4 -- the runtime link -- is addressed by complementary work on DT platforms, such as Eclipse~Ditto~\cite{kherbache2022ditto
}. It is worth noting that integrating Modelio models with such platforms is the natural next step of our research line~\cite{bucaioni2025matisse}.

Consequently, it is necessary to choose the most suitable validation framework for our ontology-based solution. Shapes Constraint Language (SHACL)~\cite{pareti2021review} can be considered as an appropriate choice, as it is provided specifically to validate Resource Description Framework (RDF) graphs, aligning perfectly with the RDF-based structure of SAREF4ENER. Through this framework, constraints can be defined and applied directly to RDF data, guaranteeing that the data complies with the logical and structural standards established by the ontology. In contrast to Object Constraint Language (OCL)~\cite{brucker2015featherweight}, which is primarily applied for UML models, SHACL can integrates seamlessly with RDF tools and also offers flexibility in defining sophisticated validation rules. Furthermore, its declarative nature guarantees the clarity and ease of maintenance of validation rules. The development of robust, ontology-based solutions in different industries (e.g., the energy sector) is supported by this conceptual framework, which enables efficient communication, seamless data integration, and interoperability across diverse platforms and technologies. It is worth noting that pySHACL can be used to implement the SHACL framework, a Python library that offers automated, constraint-based validation of RDF data, as well as to further improve data quality and interoperability~\cite{chy2024design}.

\section{Related Work}
\label{related_work}
The use of SysML as a design-time substrate for DTs has been explored along several lines. Wilking et al.~\cite{wilking2022sysml4dt,wilking2024assembly} derive the executable DT behavior from SysML diagrams for generic systems and, more recently, for smart assembly lines. Pessoa et al.~\cite{pessoa2023mbse} formalize an eight-step MBSE methodology that integrates SysML with discrete-event simulation. Ferko et al.~\cite{ferko2025aas} generate Asset Administration Shells from SysML~v2 models, bridging the gap between MBSE and the Industry~4.0 reference architecture. Related efforts apply SysML to BIM-based~\cite{keskin2022bim} and modular systems-of-systems~\cite{wagner2023sysml} DTs. The recent systematic mapping study by Lehner et al.~\cite{lehner2025mde} confirms the momentum of model-driven DT engineering, but also shows that published applications focus on manufacturing, while smart energy and renewable energy communities remain markedly underrepresented.

A parallel body of work tackles the complementary problem of semantic interoperability for smart energy systems. Beyond the original SAREF ontology~\cite{daniele2016interoperability}, alternative vocabularies, such as SARGON~\cite{haghgoo2020sargon} and more recent unified ontologies~\cite{ivanova2025unified}, have been proposed, along with alignments between SAREF and neighboring standards -- notably IFC~\cite{okonta2025aligning} and EnergyPlus~\cite{kirnapci2025ontology}. Aniakor et al.~\cite{aniakor2024semantic} provide a survey of semantic models for building energy management, and Lilliu et al.~\cite{lilliu2025flexoffers} extend SAREF4ENER with the FlexOffer concept to better support demand-response markets. These contributions are complementary to, rather than competing with, the structural SysML layer we target.

On the runtime side, Eclipse~Ditto has been investigated as a DT backbone for Industrial~IoT~\cite{kherbache2022ditto} and combined with frameworks such as Eclipse~Arrowhead~\cite{aziz2022arrowhead} and OpenTwins~\cite{robles2023opentwins}. Surveys specifically dedicated to DTs for renewable energy, microgrids, and smart grids~\cite{alshetwi2025dt
,bassey2024microgrids
} confirm a rapid rise of interest, but consistently note the absence of a systematic engineering methodology.

To the best of our knowledge, no prior work explicitly combines SysML-based MBSE with SAREF or SAREF4ENER semantics for renewable energy communities. The two communities evolve in parallel: structural/architectural models on one side; ontologies and flexibility profiles on the other. The present paper is a first step towards bridging them, in continuity with earlier work on SysML as a co-simulation and federated DT backbone~\cite{sadovykh2016sysml,bucaioni2025matisse}.

\section{Conclusion}
\label{conclusion}
We presented a first step towards a Model-Based Systems Engineering workflow for Digital Twins of Renewable Energy Communities. Starting from 
an industrially-validated REC domain model, we expressed a representative \emph{household} subset in SysML using the open-source Modelio tool, yielding two Block Definition Diagrams -- a device taxonomy and a REC community organizational view. We then identified four semantic gaps that plain SysML leaves open and sketched how the SAREF4ENER ontology could be imported as a reference package to close them, without resorting to heavy custom stereotypes. Related work confirms that combining SysML with SAREF-based semantics for smart energy Digital Twins has received the little attention so far, and we position this paper as a starting point along that line. The ongoing work aims to fill the fourth gap -- generating Eclipse~Ditto \emph{Thing} definitions based on the annotated SysML model, so that the design-time model becomes the live runtime reference, an instance of the Models@Run.Time vision applied to energy communities.

\section*{Acknowledgment}
This research work has been partially funded by the Chips Joint Undertaking through the project MATISSE (grant agreement No.\ 101140216).

\bibliographystyle{ieeetr}
\bibliography{bibliography}
\balance

\end{document}